\documentclass[article,twocolumn,showpacs,floatfix,longbibliography]{revtex4-2}
\usepackage{mathrsfs,braket}
\usepackage{amssymb, amsbsy, amsmath, latexsym, dsfont, array, layout, graphicx, mathrsfs, color, ulem, bm}
\usepackage{amsthm}
\usepackage[colorlinks=true,linkcolor=blue,anchorcolor=red,citecolor=blue, urlcolor=blue]{hyperref}

\begin{document}
\title{Topological origin of non-Hermitian skin effect in higher dimensions and uniform spectra}

\author{Haiping Hu\textsuperscript{1, 2}}\email{hhu@iphy.ac.cn}
\affiliation{\textsuperscript{1}Beijing National Laboratory for Condensed Matter Physics, Institute of Physics, Chinese Academy of Sciences, Beijing 100190, China}
\affiliation{\textsuperscript{2}School of Physical Sciences, University of Chinese Academy of Sciences, Beijing 100049, China}
\begin{abstract}
\textbf{Abstract}~The non-Hermitian skin effect is an iconic phenomenon characterized by the aggregation of eigenstates near the system boundaries in non-Hermitian systems. While extensively studied in one dimension, understanding the skin effect and extending the non-Bloch band theory to higher dimensions encounters a formidable challenge, primarily due to infinite lattice geometries or open boundary conditions. This work adopts a point-gap perspective and unveils that non-Hermitian skin effect in all spatial dimensions originates from point gaps. We introduce the concept of uniform spectra and reveal that regardless of lattice geometry, their energy spectra are universally given by the uniform spectra, even though their manifestations of skin modes may differ. Building on the uniform spectra, we demonstrate how to account for the skin effect with generic lattice cuts and establish the connections of skin modes across different geometric shapes via momentum-basis transformations. Our findings highlight the pivotal roles point gaps play, offering a unified understanding of the topological origin of non-Hermitian skin effect in all dimensions.\\


\end{abstract}
\maketitle
\textbf{1. Introduction}\\

Non-Hermitian systems have emerged as a captivating realm of exploration in recent years \cite{coll1,coll4,coll6,colladd3,nhreview,nhreview2,nhreview3}, encompassing a wide range of classical wave systems such as lossy photonic crystals \cite{photonic2,op2,op5,op8,op17,op19} or acoustic cavities \cite{ep2encircling3,epnexus,pg3,hep1,qzhang}, ultracold atoms \cite{yanbo,hkust,rosa,coldatom1}, and electronic systems hosting quasiparticles of finite lifetimes \cite{finite1}. The breaking of Hermiticity enables the appearance of complex eigenenergies, giving rise to intrinsic phenomena that lack analogues in Hermitian systems. A prototypical example is the non-Hermitian skin effect (NHSE) \cite{nhse1,nhse3,nhse4,nhse5,nhse6,nhse7,nhse8,nhse9,nhsereview,pointtopo3}, which exhibits distinct energy spectra under different boundary conditions and the localization of a large number of eigenmodes at the boundaries. Experimental observations of the NHSE have been reported in various platforms \cite{nhsee1,nhsee2,nhsee3,nhsee4}, and potential applications, such as unidirectional amplifiers \cite{nhsea1}, optical funnels \cite{nhsefunnel}, high-efficiency energy harvesting \cite{nhsea2}, and enhanced sensors \cite{nhsea3,nhsea4,nhsea5}, have been explored.

The complex energy spectra bring novel notions to non-Hermitian systems such as point gap \cite{pgclass1,pgclass2,pgclass3,pgclass4,pgclass5}. In Hermitian Hamiltonians, all eigenenergies are real, and only line gaps are allowed. However, in the non-Hermitian case, the eigenenergies may exhibit diverse patterns \cite{huknot,huep,huepna} on the complex plane, deviating from a specific reference point, i.e., a point gap. At the heart of the NHSE lies the point gap. It dictates the supersensitivity of the complex energy spectrum to the imposed boundary condition or threading an imaginary flux \cite{nhse6,nhse9,spumping}, challenging the conventional Bloch band theory. A synopsis of one-dimensional (1D) non-Hermitian band theory involves replacing the Bloch wave vector $k\rightarrow k+i\kappa$ to account for the presence of skin modes. With the localization properties of these skin modes encoded in $\kappa$, this analytical continuation bridges the non-Hermitian band structures under different boundary conditions, reinstating the usual bulk-edge correspondence.

\begin{figure}[!t]
\centering
\includegraphics[width=3.33 in]{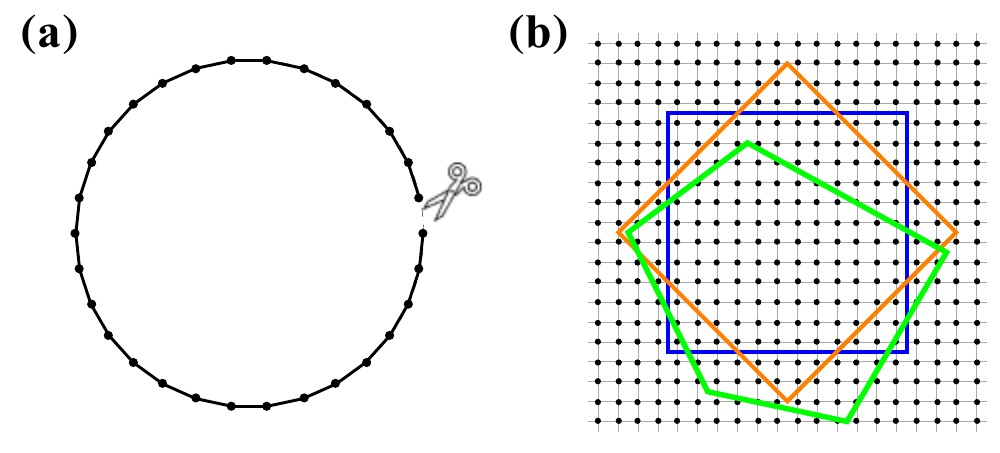}
\caption{Sketch of lattice cuts and open boundary conditions (OBC). (a) 1D case: OBC is achieved by cutting off the lattice across a single bond. (b) 2D case: Open boundary can be oriented in any direction, leading to diverse geometric shapes (in different colors) defined by the lattice cuts.}\label{fig1}
\end{figure}

Extending the non-Bloch band theory to higher dimensions, however, presents formidable challenges. Higher-dimensional systems introduce complexities due to the vast diversity of lattice geometries and open boundary conditions (OBC) (See Fig. \ref{fig1}). The lattice cuts may influence the presence and localizations of skin modes \cite{fangchen}. Moreover, the vexing numerical errors \cite{error1,toeplitzbook,error2} seem unavoidable, especially when the system size is large. Previous investigations of higher-dimensional non-Hermitian systems \cite{nhse2,james,litianyu} have mostly been conducted on a case-by-case basis. Recent advancements have shed light on the understanding of NHSE in higher dimensions. For instance, a global criterion has been proposed that relates the occurrence of NHSE to the spectral area \cite{fangchen}. An elegant formulation \cite{wz} connects the spectral density to the mathematical object Amoeba. For certain 2D models, ``dimensional transmutation" of the generalized Brillouin zone may occur \cite{jianghuilee}. Additionally, a dynamical degeneracy splitting has been introduced to characterize the anisotropic decay behaviors \cite{dynamicals1,dynamicals2}. Despite these developments, fundamental questions regarding NHSE in higher dimensions remain elusive: What is the topological origin of NHSE in higher dimensions? How are skin effects with distinct lattice cuts interrelated, and how can we account for the energy spectra and skin effects across various lattice geometries in a unified way?

In this paper, we tackle these problems by presenting a point-gap perspective of NHSE that applies to all spatial dimensions. Drawing insights from the 1D case with point gaps, we introduce the key notion of uniform spectra ($\sigma_{\rm U}$) from the perspective of inserting imaginary flux. We illustrate that the uniform spectra are lattice-cut independent and precisely describe the energy spectra for any lattice geometries. Based on the uniform spectra, we reveal that the skin modes on different lattice geometries can be treated as different ``projections" of the same entity, which are further related through momentum-basis transformations.\\

\textbf{2. Analytical continuation}\\

Let us first revisit the well-established non-Bloch band theory in 1D. The Hamiltonian on a 1D lattice is expressed as:
\begin{eqnarray}
H=\sum_{i,j=1}^L t_{i-j}c_{i}^{\dag} c_j,
\end{eqnarray}
where $c_i$ ($c_{i}^{\dag}$) represents the annihilation (creation) operator on the $i$-th site. For multi-band systems, the internal degrees of freedom (e.g., spin, sublattice) can be incorporated into the annihilation/creation operators. Through Fourier transformation into momentum space, the Hamiltonian takes the Bloch form:
\begin{eqnarray}
H(k)=\sum_{m=-p}^q t_{m} e^{i mk}.
\end{eqnarray}
Here $p\geq 0$ and $q\geq 0$ represent the hopping range to the left and right side, respectively. The energy spectra under periodic boundary condition (PBC), are given by $\sigma_{\rm PBC}=\{ H(k),~|e^{ik}|=1\}.$ The eigenmodes manifest as Bloch waves that extend throughout the entire lattice.

To describe the skin modes under OBC, the non-Bloch band theory sets in by replacing the Bloch wave vector $k\rightarrow k+i \kappa$ to incorporate the localization of the skin modes. The Bloch Hamiltonian is analytically continued to $H(\beta)$ with $\beta=e^{ik}\in\mathbb{C}$. The OBC energy spectra are extracted from the characteristic polynomial (ChP):
\begin{eqnarray}
f(E,\beta)=\det(H(\beta)-E)=\sum_{m=-p}^q t_m\beta^m-E,
\end{eqnarray}
which is a Laurent series of two complex variables $(E,\beta)$. For a given $E$, one can find $p+q$ solutions of $\beta$. Sorting them by the modulus, $|\beta_{(1)}(E)|\leq|\beta_{(2)}(E)|\leq\ldots\leq|\beta_{(p+q)}(E)|$, the OBC energy spectra are given by \cite{nhse1,nhse4}:
\begin{eqnarray}\label{obc1}
\sigma_{\rm OBC}=\{E,~| \beta_{(p)}(E)|=|\beta_{(p+1)}(E)| \}.
\end{eqnarray}
The solutions $(\beta_{(p)}(E),\beta_{(p+1)}(E))$ satisfying $|\beta_{(p)}(E)|=|\beta_{(p+1)}(E)|$ give the generalized Brillouin zone, with their moduli determining the localization lengths of the skin modes.

Now, let us delve into the $d$D case with Bloch Hamiltonian $H(\bm k)=H(k_1,k_2,...,k_d)$. The first complexity we encounter is the selection of boundary cuts. There are infinitely many lattice geometries, as depicted in Fig. \ref{fig1}b. There is no guarantee that different geometric shapes yield the same OBC spectra \cite{fangchen}, nor a prior rule to connect skin modes on different geometries. Furthermore, it is not clear how to handle irregular shapes, such as the pentagon shown in Fig. \ref{fig1}b. So, how should we address these diverse situations within the unique Bloch Hamiltonian $H(\bm k)$? The second difficulty arises from the algebraic aspects. Let us formally replace the Bloch wave vector with its non-Bloch form for each lattice momentum: $k_j\rightarrow k_j+i\kappa_j$. Denoting $\beta_j=e^{i k_j}$ and analytically continuing it to the entire complex plane with Hamiltonian $H(\beta_1,\beta_2,...,\beta_d)$, the eigenspectra of the system are determined by the ChP:
\begin{eqnarray}
f(E,\beta_1,\beta_2,...,\beta_d)=\det(H(\beta_1,\beta_2,...,\beta_d)-E).
\end{eqnarray}
The ChP is an algebraic equation with $1+d$ complex variables. Note that, as a Laurent series for each individual $\beta_j$, its maximum order may differ (unlike $p,q$ in the 1D case). For a given $E$, there exist infinitely many solutions of $(\beta_1,\beta_2,...,\beta_d)$, and it seems hard to generalize the condition $|\beta_{(p)}(E)|=|\beta_{(p+1)}(E)|$ \cite{wz}.\\

\textbf{3. Results}\\

\textit{3.1 Point gap and uniform spectra}\\

\begin{figure}[!t]
\centering
\includegraphics[width=3.3 in]{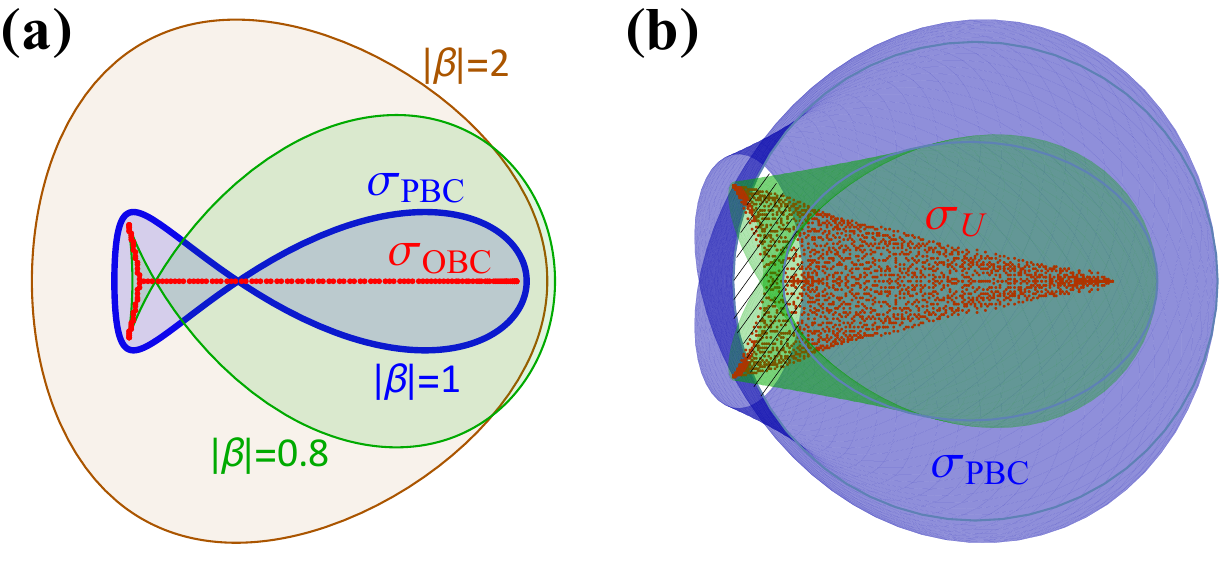}
\caption{Different types of energy spectra. (a) The 1D case with model $H(\beta)=\beta+\beta^{-1}+\frac{2}{5}\beta^{-2}$. It displays the PBC spectra $\sigma_{\rm PBC}$ (blue loop), OBC spectra $\sigma_{\rm OBC}$ (red arcs), and the rescaled spectra $Sp(0.8)$ (green region), $Sp(1)$ (blue region), $Sp(2)$ (orange region). (b) The 2D case with model $H(\beta_1,\beta_2)=\beta_1+\frac{1}{2}\beta_1^{-1}+\beta_2+\frac{1}{2}\beta_2^{-1}+2\beta_1\beta_2$. It displays the PBC spectra $\sigma_{\rm PBC}$ (blue region), rescaled spectra $Sp(1,1)$ (blue region and shading region) and $Sp(0.8,0.8)$ (green region), and the uniform spectra $\sigma_{\rm U}$ (red region).}\label{fig2}
\end{figure}

Our strategy is to adopt a point-gap perspective and examine the spectral evolution by introducing an imaginary flux. In the context of 1D NHSE, the PBC spectra form closed loops on the complex energy plane. The OBC spectra are entirely contained within these spectral loops, as shown in Fig. \ref{fig2}a. By introducing an imaginary flux, the PBC spectral loop is unstable and undergoes deformation through a minimal coupling $k\rightarrow k-i r$. It corresponds to an imaginary gauge transformation on the system: $H(k)\rightarrow H(k-i r)$. This transformation does not alter the OBC spectra, as it is implemented as a similarity transformation at the matrix level. Remarkably, the OBC spectra $\sigma_{\rm OBC}$ remain enclosed within the deformed spectral loops for all possible flux strengths \cite{nhse6}:
\begin{eqnarray}\label{point}
\sigma_{\rm OBC}\subseteq\mathop{\bigcap}\limits_{|\beta|\in(0,\infty)} Sp(|\beta|).
\end{eqnarray}
Here, $Sp(|\beta|)$ denotes the region enclosed by the deformed spectral loop (including the loop itself). The term ``enclosed" refers to a non-zero spectral winding number with respect to the fixed $|\beta|$:
\begin{eqnarray}
w(E)=\frac{1}{2\pi i}\oint_{0}^{2\pi} \partial_{\theta}\log f(E,|\beta|e^{i\theta}).
\end{eqnarray}
In fact, Eq. (\ref{point}) can be strengthened to be an identity (as a special case of the theorem below). Physically, the spectral region $Sp(|\beta|=1)$ corresponds to the energy spectra under half-infinite boundary conditions and is bounded by $\sigma_{\rm PBC}$ \cite{nhse6}. It is evident that the 1D NHSE originates from the existence of point gap: the point gap with a nonzero $w(E)$ inside the PBC spectra ($|\beta|=1$) would eventually be intersected by some deformed loops corresponding to some flux strengths. By taking the intersections in Eq. (\ref{point}), the OBC spectra must differ from the PBC spectra.

This perspective can be readily extended to higher dimensions. By inserting $d$ imaginary fluxes along each direction, we define:
\begin{eqnarray}\label{sp}
Sp(|\beta_1|,|\beta_2|,...,|\beta_d|):=\{E,~\sum_{j=1}^d|w_j(E)|\neq 0\},\label{pg}\\
\textrm{with}~w_j(E)=\frac{1}{2\pi i}\oint_{0}^{2\pi} \partial_{\theta_j}\log f(E,\beta_1,\beta_2,...,\beta_d). \label{winding}
\end{eqnarray}
Here, $w_j(E)$ is the winding number along the $j$-th direction. The condition $\sum_{j=1}^d|w_j|\neq 0$ captures the intuitive notion that the NHSE can arise from a point gap in any direction. For a fixed set of rescaling factors $(|\beta_1|,|\beta_2|,...,|\beta_d|)$, $Sp(|\beta_1|,|\beta_2|,...,|\beta_d|)$ describes the spectral region where at least one of the $d$ directions has nonzero spectral winding or point gap. Similar to the 1D case, $Sp(|\beta_1|,|\beta_2|,...,|\beta_d|)$ has a geometric interpretation, encompassing both the rescaled Bloch spectra and the enclosed region. To visualize, let us consider a 2D model with Hamiltonian $H(\beta_1,\beta_2)=\beta_1+\frac{1}{2}\beta_1^{-1}+\beta_2+\frac{1}{2}\beta_2^{-1}+2\beta_1\beta_2$. As shown in Fig. \ref{fig2}b, $Sp(1,1)$ contains both the PBC spectra $\sigma_{\rm PBC}=H(|\beta_1|=|\beta_2|=1)$ (in blue) and the smaller shaded region inside with nonzero winding numbers $w_1=w_2=1$. For another rescaling $|\beta_1|=|\beta_2|=0.8$, we can see the OBC spectra are also inside $Sp(0.8,0.8)$. With varying the rescaling factors, we define the uniform spectra $\sigma_{\rm U}$ as the intersection of all $Sp$'s:
\begin{eqnarray}\label{obcs}
\sigma_{\rm U}=\mathop{\bigcap}\limits_{|\beta_j|\in(0,\infty), i=1,2,...,d} Sp(|\beta_1|,|\beta_2|,...,|\beta_d|).
\end{eqnarray}
Technically, the OBC energy spectra are collapsed from these rescaled spectra till no spectral winding along any direction exists. This is realized by taking the spectral intersections in Eq. (\ref{obcs}), which eliminates all possible spectral windings. The multiple-winding number description provides a possible route for a systematic classification of NHSE in higher dimensions. For example in the 2D case of geometry-dependent skin effect (GDSE) \cite{fangchen}, one can find two directions with no spectral windings. While for the normal case of corner NHSE, the winding number along the two lattice-cut directions are nonzero. \\

\textit{3.2 Geometry-independence}\\

We aim to relate the uniform spectra $\sigma_{\rm U}$ with various lattice geometries. So far, we have not yet specified any particular lattice geometry. To facilitate our discussion, we define regular lattice cuts as geometric shapes in $d$D that can be labeled by $d$ independent lattice momenta, e.g., parallelograms in 2D and parallelepipeds in 3D. Otherwise, we refer to shapes that do not meet this criterion as irregular lattice cuts.\\

We have the first conclusion that any regular lattice cut yields the same uniform spectra $\sigma_{\rm U}$.  The specific basis chosen corresponds to a definite regular lattice cut in our definition. Note that in our analysis, the representation of the ChP $f(E,\beta_1,\beta_2,...,\beta_d)$ via analytical continuation $\beta_j=e^{ik_j}$ has prescribed a momentum basis $(\hat{k}_1,\hat{k}_2,...,\hat{k}_d)$. We can associate this basis with a particular lattice cut. Specifically, the choice of momenta $(k_x,k_y,k_z)$ corresponds to the standard cubic cut along the $(\hat{x},\hat{y},\hat{z})$ directions. The lattice momenta associated with two different regular shapes are related via a transformation: $\bm k' = S \bm k$. For instance, a square-shaped lattice and a diamond-shaped lattice (See Fig. \ref{fig1}b) are related by transformation:
\begin{eqnarray*}
(k_1',k_2')^T=\frac{1}{\sqrt{2}}\left(\begin{array}{cc}
 1 & 1 \\
1 & -1\\
\end{array}\right)
(k_1,k_2)^T.
\end{eqnarray*}
Importantly, $\sigma_{\rm U}$ does not depend on such transformations. A transformation $S$ would induce a shift in the rescaling factors. For instance, a 2D Hamiltonian $\beta_1+2\beta_2$ transforms into $\beta_1'\beta_2'+2\beta_1'(\beta_2')^{-1}$ with the aforementioned transformation. However, our definition of $\sigma_{\rm U}$ in Eq. (\ref{obcs}) contains all possible rescaling factors. It is evident that sweeping all the rescalings $(|\beta_1|,|\beta_2|)$ is equivalent to sweeping all the rescalings $(|\beta_1'|,|\beta_2'|)$. Additionally, the transformation is nonsingular ($\det S\neq0$) and preserves the condition $\sum_j |w_j(E)|\neq 0$. By taking the intersections in Eq. (\ref{obcs}), $\sigma_{\rm U}$ remains unchanged.

We have the second conclusion that
\begin{eqnarray}\label{spequality}
\sigma_{\rm U}=\sigma_{\textrm{Amoeba}}.
\end{eqnarray}
Here $\sigma_{\textrm{Amoeba}}$ is another type of non-Bloch spectra obtained from the mathematical object Amoeba \cite{wz}. $\sigma_{\textrm{Amoeba}}$ is also geometry-irrelevant. A pedagogical introduction to Amoeba formulation and the detailed proof of Eq. (\ref{spequality}) can be found in the appendix. The key is to establish the connection between the central hole in Amoeba and the point gap in Eq. (\ref{pg}). With a prescribed basis $\bm k=(k_1,k_2,...,k_d)$, Ref. \cite{wz} further reveals that the OBC spectra [We note the subtlety of the OBC energy spectra associated with different lattice cuts. Here the OBC spectra represent the stable components in the presence of small perturbations.] tend to the Amoeba spectra $\sigma_{\textrm{Amoeba}}$ through Szeg\"o's strong limit theorem \cite{szego1,szego21,szego22,szego23,szego5} and logarithmic potential theory \cite{logp}. Thus the uniform spectra $\sigma_{\rm U}$ correspond to the non-Bloch spectra in the thermodynamic limit.\\

\textit{3.3 NHSE on generic lattice geometries}\\

We now shift our focus to generic lattice shapes and establish a connection between NHSE on different geometries. Drawing insights from the above discussions that the spectra on any lattice cuts uniformly tend to $\sigma_{\rm U}$, the skin modes on seemingly distinct lattices can be regarded as different ``projections" of the same entity. This observation elucidates why skin modes exhibit varying degrees of localization on different lattice shapes. Formally, let us denote the lattice momenta associated with two different cuts as $\bm k=(k_1,k_2,...,k_d)$ and $\bm k'=(k'_1,k'_2,...,k'_d)$, related through $(k'_1,k'_2,...,k'_d)=S(k_1,k_2,...,k_d)$. Considering the equal status of the inverse localization length (ILL) and lattice momentum in the analytical continuation $\beta_j=|\beta_j|e^{i k_j}=e^{i k_j+\log|\beta_j|}$, we obtain that the ILL obeys the same transformation:
\begin{eqnarray}\label{trans}
\log|\bm \beta'|=S\log|\bm \beta|,
\end{eqnarray}
with $\log|\bm \beta|=(\log|\beta_1|,\log|\beta_2|, ..., \log|\beta_d|)$ and $\log|\bm \beta'|=(\log|\beta'_1|,\log|\beta'_2|, ..., \log|\beta'_d|)$.
\begin{figure}[!t]
\centering
\includegraphics[width=3.33 in]{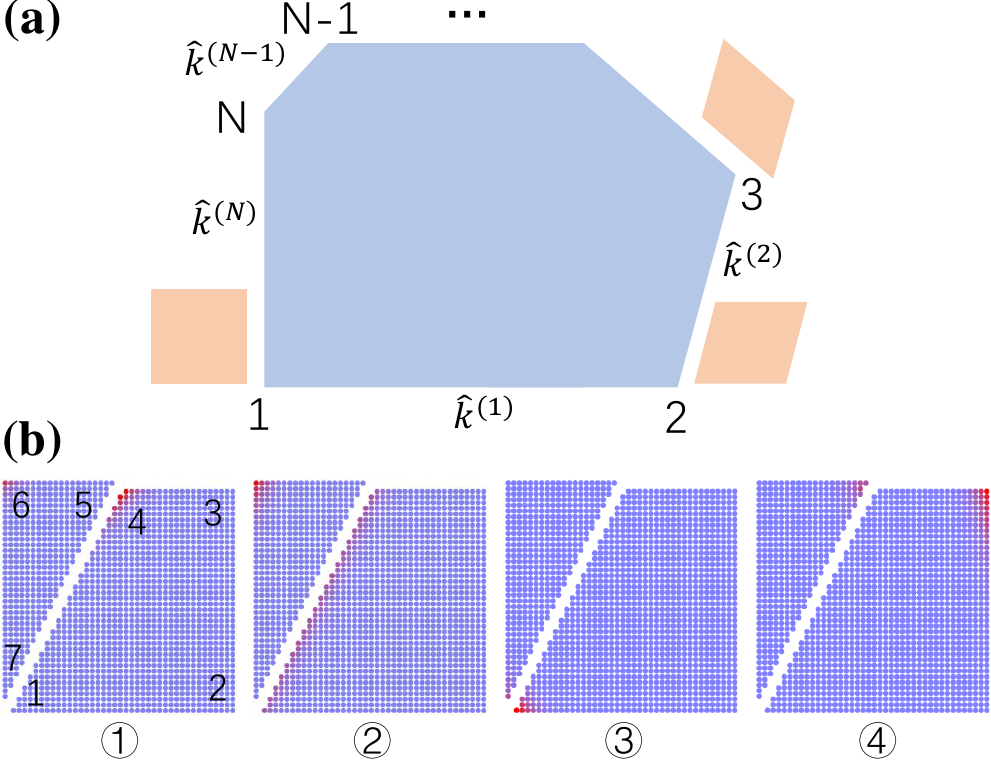}
\caption{Analysis of skin modes on generic lattice geometries. (a) Sketch of the decomposition of an $N$-polygon into $N$ simple parallelograms at the corners. Each parallelogram is formed by two neighboring cuts of the $N$-polygon. (b) Density distributions of the eigenstates for the 2D Hatano-Nelson model. The lattice cut is along the $(\hat{x}+2\hat{y})$ direction. The parameters $(t_1,t_{-1},r_1,r_{-1})$ for \textcircled{1}\textcircled{2}\textcircled{3}\textcircled{4} are $(1,2,2,1); (1,2,2,\sqrt2); (1,2,2,3); (1,1/2,2,5/3)$, respectively.}\label{fig3}
\end{figure}

For arbitrary lattice shapes, the strategy is to adopt a local perspective and decompose complex shapes, such as irregular polygons (2D) and polyhedra (3D), into manageable components. For example in 2D, an $N$-polygon has $N$ boundaries and each edge is associated with a specific cut direction denoted as $\hat{k}^{(1)}, \hat{k}^{(2)}, ..., \hat{k}^{(N)}$, as sketched in Fig. \ref{fig3}a. The corners where two neighboring boundaries meet are identified by pairs of cut directions ($d$ cut directions in $d$D), namely $(\hat{k}^{(j)}, \hat{k}^{(j+1)}), (j=1,2,...,N)$ with $\hat{k}^{(N+1)}=\hat{k}^{(1)}$. To explore the skin modes on such complicated shapes, we need to consider $N$ regular parallelograms. Taking the thermodynamic limit smears out the finite-size effects, and the skin modes on a generic geometry arise from the projections onto the parallelogram at each corner.

We illustrate the local perspective and relations of skin modes in the 2D Hatano-Nelson model with Hamiltonian $H(k_x,k_y)=t_1 e^{i k_x}+t_{-1} e^{-i k_x}+r_1 e^{i k_y}+r_{-1} e^{-i k_y}$. It has non-reciprocal hoppings along both $x$ and $y$ directions. In this primitive basis, the ILL is given by $(\log\sqrt{|t_1/t_{-1}|},\log\sqrt{|r_1/r_{-1}|})$. With a cut along the $(1,2)$ direction as depicted in Fig. \ref{fig3}b, the seven corners are grouped into three types: corners 2, 3, 6; corners 1, 4, 5; and corner 7. The basis transformations are
\begin{eqnarray*}
S_{1,4,5}=\left(\begin{array}{cc}
1 & 0 \\
\frac{1}{\sqrt{5}} & \frac{2}{\sqrt{5}}
\end{array}\right);~~~S_7=\left(\begin{array}{cc}
\frac{1}{\sqrt{5}} & \frac{2}{\sqrt{5}}\\
0 & 1
\end{array}\right).
\end{eqnarray*}
With the chosen parameters, the ILL relevant for corner 2, 3, 6 (no transformation is needed) are $\textcircled{1}:\frac{1}{2}(-\log2,\log2)$, $\textcircled{2}:\frac{1}{2}(-\log2,\frac{\log2}{2})$, $\textcircled{3}:\frac{1}{2}(-\log2,\log\frac{2}{3})$, and $\textcircled{4}:\frac{1}{2}(\log2,\log\frac{6}{5})$ for the four panels in Fig. \ref{fig3}. The sign inside the brackets refers to the localization direction. It dictates the emergence of skin modes at corner 6 in \textcircled{1}\textcircled{2} (the first/second element in ILL is negative/positive) and corner 3 in \textcircled{4} (the first/second element in ILL is positive/positive). Similarly, for corner 1, 4, 5, the transformed ILL in \textcircled{1}\textcircled{2}\textcircled{3}\textcircled{4} are $\frac{1}{2}(-\log2,\frac{1}{\sqrt 5}\log2)$, $\frac{1}{2}(-\log2,0)$, $\frac{1}{2}(-\log 2,-\frac{1}{\sqrt 5}\log 2+\frac{2}{\sqrt 5}\log\frac{2}{3})$, and $\frac{1}{2}(\log2,\frac{1}{\sqrt 5}\log2+\frac{2}{\sqrt 5}\log\frac{6}{5})$, consistent with the presence of skin modes at corner 4 in \textcircled{1}, corner 1 in \textcircled{3}, and corner 5 in \textcircled{4}. Notably in \textcircled{2}, the skin modes extend along the cut due to the vanishing of the second component in the ILL. One can also take corner 7 with the transformation $S_7$ to check the presence of skin modes in \textcircled{3}.\\

\begin{figure*}[!t]
\centering
\includegraphics[width=0.75\textwidth]{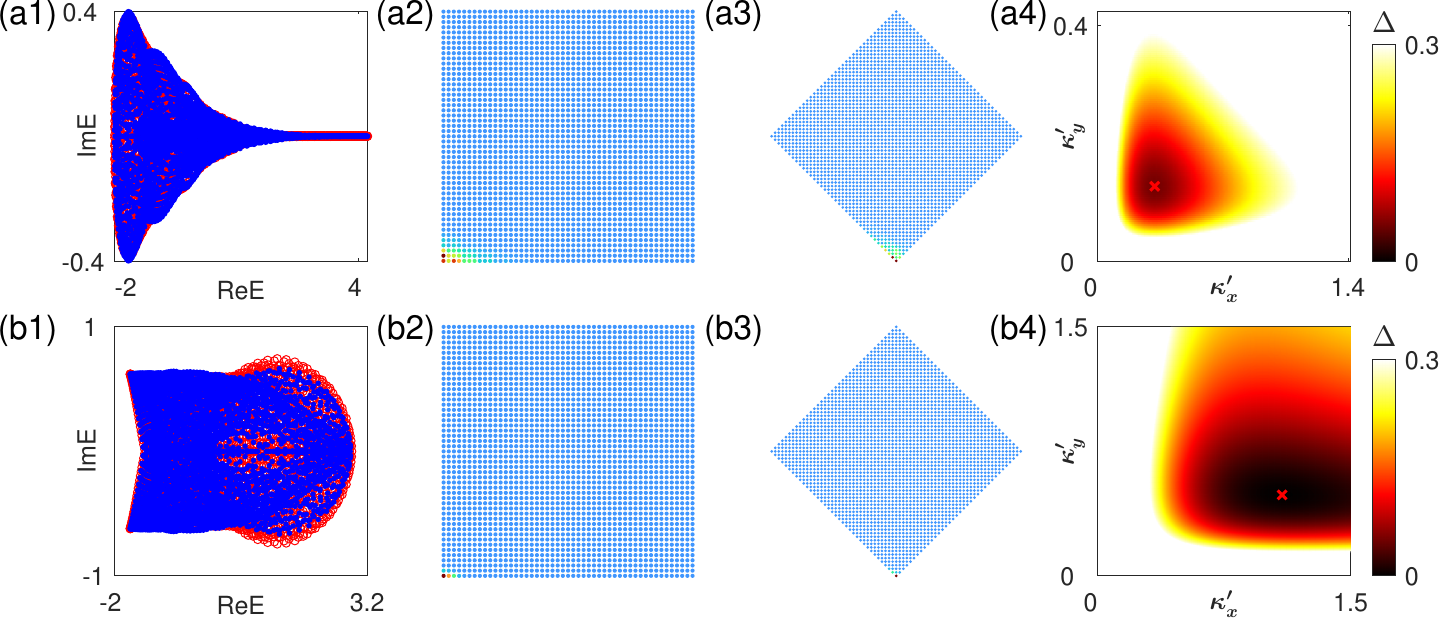}
\caption{Skin modes on different lattice geometries. The model is taken as $H(\beta_x,\beta_y)=1.2\beta_x+0.8\beta_x^{-1}+1.2\beta_y+0.4\beta_y^{-1}+0.2(\beta_x+\beta_x^{-1})(\beta_y+\beta_y^{-1})$ for (a1--a4) and $H(\beta_x,\beta_y)=\beta_x+0.5\beta_x^{-1}+\beta_y+0.2\beta_y^{-1}+2\beta_x\beta_y$ for (b1)--(b4). (a1), (b1) Energy spectra for square (red dots) and rhombus (blue dots) shaped lattices. (a2), (a3), (b2), (b3) Spatial distributions of the eigenmodes near $E=3$ and $E=1.3+0.25i$, respectively. (a4), (b4) Fitting deviations $\Delta(\kappa_x',\kappa_y')$ as described by Eq. (\ref{fitt}).}\label{fig4}
\end{figure*}
\textit{3.4 More examples}\\

The 2D Hatano-Nelson model discussed earlier can be transformed into a Hermitian lattice model through a similarity transformation. In this subsection, we consider two more general 2D models where the $x$ and $y$ directions are ``entangled" together. The first model is give by $H(\beta_x,\beta_y)=1.2\beta_x+0.8\beta_x^{-1}+1.2\beta_y+0.4\beta_y^{-1}+0.2(\beta_x+\beta_x^{-1})(\beta_y+\beta_y^{-1})$; and the second model is given by $H(\beta_x,\beta_y)=\beta_x+0.5\beta_x^{-1}+\beta_y+0.2\beta_y^{-1}+2\beta_x\beta_y$. For these two models, we investigate skin modes on two different shapes, namely, square and rhombus cuts. Their energy spectra are shown in Fig. \ref{fig4}(a1)and (b1), respectively. It can be observed that for either model, the two geometries exhibit very similar energy spectra and converge to the uniform spectra in the thermodynamic limit. In realistic calculations, a small onsite disorder may be added to stabilize the numerics.

To verify the transformation relationships of skin modes under different geometries, we consider a small region in the complex plane near $E=3$ (in Fig. \ref{fig4}(a1)) and $E=1.3+0.25i$ (in Fig. \ref{fig4}(b1)), and plot the spatial distributions of the associated eigenstates on the lattice in Fig. \ref{fig4}(a2), (a3), (b2) and (b3). We observe that these skin modes are all localized at a specific corner. To extract their localization lengths, we fit these eigenstates with exponential functions and then average over the selected eigenstates. With the square geometry, the fitting function is taken as $|\psi_{fit}(x,y)|\sim e^{-\kappa_x x-\kappa_y y}$, where $(x,y)$ label the lattice sites along $x$ and $y$ directions. For the two models, the inverse localization lengths are fitted to be $\kappa_{\rm square}=(0.135,0.313)$ and $\kappa_{\rm square}=(0.425,1.105)$, respectively. With the rhombus geometry, the fitting function is taken as $|\psi_{\rm fit}(x',y')|\sim e^{-\kappa_x' x'-\kappa_y' y'}$, where $(x',y')$ label the layer indices along the $(1,1)$ and $(1,-1)$ directions. In Fig. \ref{fig4}(a4) and (b4), we plot the disparity between numerically obtained eigenstates and the fitting functions with respect to the fitting parameters $(\kappa_x',\kappa_y')$:
\begin{eqnarray}\label{fitt}
\Delta(\kappa_x',\kappa_y')=\frac{1}{N}\sum_{x',y'}|\psi(x',y')- e^{-\kappa_x' x'-\kappa_y' y'}|. \end{eqnarray}
Here $N$ is the number of total lattice sites. It is evident that for the two models, the optimal fittings are $\kappa_{\rm rhombus}=(0.318,0.127)$ and $(1.093,0.488)$. The basis transformation relations $\kappa_{\rm rhombus}=S \kappa_{\rm square}$ are approximately satisfied (the deviation from exact values is due to finite-size effect and fitting errors) for both models.\\

\textbf{4. Conclusion and discussion}\\

In conclusion, we expand the point-gap perspective of 1D NHSE and present a geometry-irrelevant formulation of non-Bloch spectra across all dimensions. We show that the uniform spectra $\sigma_{\rm U}$ are invariant under basis transformations and equivalent to the spectra from the Amoeba formulation. Based on the uniform spectra, we demonstrate how to deal with the skin effect on regular lattice geometries. We further reveal that the skin modes associated with different lattice cuts are related through basis transformations and provide insights into handling arbitrary lattice geometries from a local viewpoint.

Our work indicates that NHSE in all dimensions originates from the inherent point-gap topology. It provides guidelines for manipulating skin modes via the appropriate tailoring of lattices in band engineering. We note that our formulation only addresses the macroscopic ``skin" modes, with their number scaling as $O(L^d)$. Other types of skin modes, e.g., higher-order skin modes \cite{honhse,honhse2} and hybrid skin modes \cite{nhse2d1,nhse2d2,nhseweyl,2ndorder1,2ndorder2,nhse2d3}, which have a band-topology origin and scale as $O(L^j)$ with $j<d$, fall outside the scope of this paper. Additionally, we acknowledge the unresolved issue of the geometry-dependent skin effect \cite{fangchen}, where different lattice cuts exhibit varying densities of states. And with fine-tuned boundaries, the skin modes may completely vanish, thereby violating the transformation relation Eq. (\ref{trans}). In fact, for the case of geometry-dependent skin effect (and some other cases with high spectral degeneracies), the spectral structures are highly unstable against perturbations. For these cases, adding random disorder would drive the energy spectra towards the uniform spectra or Amoeba spectra. A pertinent question is: Can we establish a generic classification of NHSE based on such geometry dependence, the multiple winding numbers, and spectral instability? Moreover, in the case of irregular geometries, while our decomposition procedure proves effective for many models in numerical calculations, can some rigorous argument be performed for its validity? Other interesting questions include the stability of the uniform spectra and skin modes in different geometries. Exploring these special cases remains an intriguing topic for future study.\\

\textbf{Conflict of interest}\\

The author declares that he has no conflict of interest.\\

\textbf{Acknowledgments}\\

This work was supported by the National Key Research and Development Program of China (2023YFA1406704 and  2022YFA1405800) and the start-up grant of IOP-CAS. \\

\textbf{Appendix A. Supplementary materials}\\

See supplementary material for the proof of the equivalence between the uniform spectra and the Amoeba spectra.

\clearpage
\appendix
\renewcommand{\thefigure}{S\arabic{figure}}
\setcounter{figure}{0}
\pagebreak
\widetext
\begin{center}
\textbf{\large  Supplemental material}
\end{center}

\begin{figure}[!h]
\centering
\includegraphics[width=3.8 in]{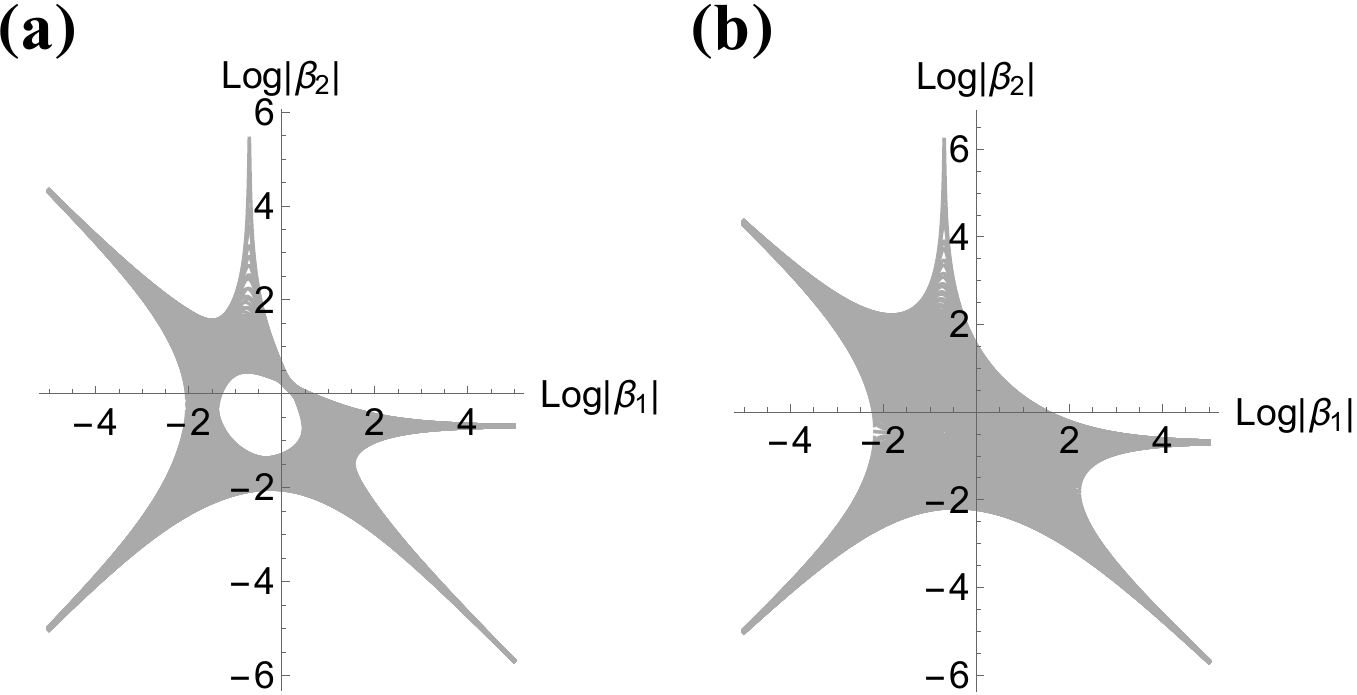}
\caption{Amoeba for the 2D model $H(\beta_1,\beta_2)=\beta_1+\frac{1}{2}\beta_1^{-1}+\beta_2+\frac{1}{2}\beta_2^{-1}+2\beta_1\beta_2$ with reference energy chosen as (a) $E_1=-2+2i$ (outside the OBC spectra region); and (b) $E_2=2+0.2i$ (inside the OBC spectra region).}\label{fig6}
\end{figure}

In this supplemental material, we provide a pedagogical introduction to Amoeba spectra and prove the equivalence relation $\sigma_U=\sigma_{\text{Amoeba}}$ presented in the main text. Here, $\sigma_U$ refers to the uniform spectra defined in Eq. (\ref{obcs}). To begin, we delve into the algebraic aspects of the characteristic polynomial (ChP). The ChP is expressed as: 
\begin{eqnarray} f(E,\beta_1,\beta_2,...,\beta_d)=\det(H(\beta_1,\beta_2,...,\beta_d)-E), 
\end{eqnarray} 
which involves $1+d$ complex variables. For a given energy $E$, there exist infinite solutions for $(|\beta_1|,|\beta_2|,...,|\beta_d|)$. These solutions can be plotted on a $d$-dimensional ($d$D) coordinate system, forming a shape known as Amoeba \cite{amoebamath}. In 1D, it is well-known that when the reference energy $E$ belongs to $\sigma_{OBC}$, the two ``middle" roots have the same modulus, i.e., $|\beta_{(p)}|=|\beta_{(p+1)}|$. To get an intuition about the Amoeba, let us consider the 2D example in the main text [see Fig. \ref{fig2}(b)]. The corresponding Hamiltonian is given by $H(\beta_1,\beta_2)=\beta_1+\frac{1}{2}\beta_1^{-1}+\beta_2+\frac{1}{2}\beta_2^{-1}+2\beta_1\beta_2$. We consider two reference energies: $E_1=-2+2i$ (outside the OBC spectra region) and $E_2=2+0.2i$ (inside the OBC spectra). The corresponding Amoeba are depicted in Fig. \ref{fig6}(a) and (b), respectively. It is evident that there is a central hole on the $(\log|\beta_1|,\log|\beta_2|)$ plane for the former; whereas there is no such central hole for the latter.

In the example above, the existence of a central hole in the Amoeba appears to be related to the OBC spectra. Interestingly, based on the shape of the Amoeba, a type of spectra is defined \cite{wz}, dubbed Amoeba spectra. Here, we reinstate this definition:
\begin{eqnarray} \label{amoebas}
\sigma_{\text{Amoeba}}:=\{\textrm{Any}~E~\textrm{without a central hole}.\}
\end{eqnarray}
In this context, a central hole refers to a region where the winding numbers along all directions become zero when the rescaling factors are taken inside that hole. The winding numbers are those defined in Eq. (\ref{winding}) of the main text. Now we are ready to prove the third step of the theorem: $\sigma_U=\sigma_{\text{Amoeba}}$.\\

\noindent\textit{Proof}: We start with the 1D case, where the Amoeba spectra are given by Eq. (\ref{obc1}). (1) First, let us consider $E\notin \sigma_{\text{Amoeba}}$, which means $|\beta_{(p)}(E)|\neq |\beta_{(p+1)}(E)|$. In this case, we can choose an intermediate value: $|\beta_{(p)}|<\xi<|\beta_{(p+1)}|$. Consequently, there are exactly $p$ zeros of the ChP inside the circle of radius $\xi$. From the complex analysis, the winding number for the reference energy $E$ with the rescaling factor $\xi$ is equal to the difference between the number of zeros and poles (which is $p$) inside the circle of radius $\xi$: $w=n_{\text{zeros}}-p$. Thus, we have $w=0$. According to our definition of $\sigma_U$ in Eq. (\ref{obcs}), $E\notin\sigma_U$, because $\sigma_U$ represents the intersection of all $Sp(|\beta|)$'s. Therefore, $\sigma_U\subseteq\sigma_{\text{Amoeba}}$. (2) Second, let us consider $E\notin\sigma_U$, meaning that there exists some $\xi$ such that $E\notin Sp(\xi)$, or in other words, $w=0=n_{\text{zeros}}-p$ or $n_{\text{zeros}}=p$. That is there are exactly $p$ zeros inside the circle of radius $\xi$. Consequently, $|\beta_{(p)}(E)|\neq|\beta_{(p+1)}(E)|$, implying $E\notin\sigma_{\text{Amoeba}}$. Hence, $\sigma_{\text{Amoeba}}\subseteq\sigma_U$. Combining (1) and (2), we can conclude that $\sigma_U=\sigma_{\text{Amoeba}}$ in 1D.

Now, let us extend the proof to higher dimensions. (1) First, consider $E\notin \sigma_{\text{Amoeba}}$. In this case, there exists a central hole, and one can find $(\xi_1,\xi_2,...,\xi_d)$ with $w_1=w_2=...=w_d=0$. According to our definition, $E\notin Sp(\xi_1,\xi_2,...,\xi_d)\subseteq\sigma_U$. Hence, $\sigma_U\subseteq\sigma_{\text{Amoeba}}$. (2) Second, let us take an $E\notin\sigma_U$, that is, there exist some $(\xi_1,\xi_2,...,\xi_d)$ such that $E\notin Sp(\xi_1,\xi_2,...,\xi_d)$; Note that $Sp$ contains any $E$ with nonzero windings. Put in another way, $w_1(E)=w_2(E)=...=w_d(E)=0$. Thus $(\xi_1,\xi_2,...,\xi_d)$ forms a central hole of Amoeba and according to the definition Eq. (\ref{amoebas}), $E\notin\sigma_{\textrm{Amoeba}}$. Combining (1) and (2) and the 1D case, we have $\sigma_U=\sigma_{\textrm{Amoeba}}$ for all dimensions. \qed

\end{document}